\documentclass[preprint,aps,superscriptaddress]{revtex4}
\usepackage{graphicx,epsfig,amssymb,amsmath,array}
\usepackage{relsize,placeins,float}
\usepackage[usenames]{color}

\begin{document}

\title{Fractional damping enhances chaos in the nonlinear Helmholtz oscillator}
\author{Adolfo Ortiz}
 \affiliation{Centro de Investigaci\'{o}n en Micro y Nanotecnolog\'{i}a, Facultad de Ingenier\'{i}a, Universidad Veracruzana, Calz. Ruiz Cortinez 455, CP 94294, Boca del R\'{i}o, Veracruz, M\'{e}xico}

\author{Jianhua Yang}
\affiliation{School of Mechatronic Engineering, China University of Mining and Technology, \\
Xuzhou 221116, P. R. China}

\email[]{mattiatommaso.coccolo@urjc.es}\author{Mattia Coccolo}
\affiliation{Nonlinear Dynamics, Chaos and Complex Systems Group,
Departamento de F\'{i}sica , Universidad Rey Juan Carlos,
Tulip\'{a}n s/n, 28933 M\'{o}stoles, Madrid, Spain}

\author{Jes\'{u}s M. Seoane}
 \affiliation{Nonlinear Dynamics,
Chaos and Complex Systems Group, Departamento de F\'{i}sica ,
Universidad Rey Juan Carlos, Tulip\'{a}n s/n, 28933 M\'{o}stoles,
Madrid, Spain}

\author{Miguel A.F. Sanju\'{a}n}
\affiliation{Nonlinear Dynamics, Chaos and Complex Systems Group,
Departamento de F\'{i}sica , Universidad Rey Juan Carlos,
Tulip\'{a}n s/n, 28933 M\'{o}stoles, Madrid, Spain}

\date{\today}

\begin{abstract}

The main purpose of this paper is to study both the underdamped and
the overdamped dynamics of the nonlinear Helmholtz oscillator with a
fractional order damping. For that purpose, we use the Gr\"unwald-Letnikov
fractional derivative algorithm in order to get the numerical
simulations. Here, we investigate the effect of taking the fractional
derivative in the dissipative term in function of the parameter
$\alpha$. Our main findings show that the trajectories can remain
inside the well or can escape from it depending on $\alpha$ which
plays the role of a control parameter. Besides, the parameter
$\alpha$ is also relevant for the creation or destruction of chaotic
motions. On the other hand, the study of the escape times of the
particles from the well, as a result of variations of the initial
conditions and the undergoing force $F$, is reported by the use of
visualization techniques such as basins of attraction and
bifurcation diagrams, showing a good agreement with previous
results. Finally, the study of the escape times versus the
fractional parameter $\alpha$ shows an exponential decay which goes
to zero when $\alpha$ is larger than one. All the results have been
carried out for weak damping where chaotic motions can take place in
the non-fractional case and also for a stronger damping (overdamped
case), where the influence of the fractional term plays a crucial
role to enhance chaotic motions. We expect that these results can be
of interest in the field of fractional calculus and its
applications.

\end{abstract}

\maketitle

\section{Introduction}

Fractional calculus is an area of the mathematical analysis. It
studies the possibilities of taking real or even complex numbers as
orders of the integral and derivatives on a known or unknown
function. Such operators are rather useful in science and
engineering. Also, fractional differential operators involve defined
integrals over a time domain, this poses significant memory effects
as shown in Ref.~\cite{book:Yang} . For these reasons, the
fractional calculus has gained much attention and relevance in the
past few years due to its applications to several research fields
such as control systems, nonlinear oscillators, potential fields,
diffusion problems, viscoelasticity and rheology, synchronization,
thermodynamics, biology among
others~\cite{Wang:Huang,Ezz,Baleanu,article:yang,AMINA:2019,SHUO:2020,HeJi,ionescu}.
In the case of nonlinear oscillators, the research has been very
extensive with interesting results, especially in the study of the
effect of introducing a fractional derivative in the damping
term~\cite{NIU:2019}. In the last reference, the Authors carried out
the study of chaos detection in the Duffing oscillator by using the
Melnikov Method. On the other hand, the detection of chaos in
fractional systems by using SALI has been developed in
Ref.~\cite{HE:2019}. Furthermore, the fractional Duffing equation in
the presence of nonharmonic external perturbations has been also
studied with detail in Ref.~\cite{JGV:2013}. Recent and very
complete reviews on the current research and real applications of
fractional calculus in science and engineering can be found in
Refs.~\cite{REVIEW:APLICATIONS,VAZQUEZ:2020,Yang,Ouannas,Wang,antegana}.

In our present work, due to the fact that the dissipation plays an
important role in the evolution of the systems, we focus our
interest in analyzing the effect of a fractional damping on  a
nonlinear oscillator dynamics. In fact, we modify the equation of
motion of the Helmholtz oscillator, which is being used in many
physical problems, in order to introduce a fractional differential
operator in the damping term instead of the classical derivatives.
Then, we focus our attention on the effect of the fractional
operator on the dynamics of the system by
studying the behavior of the trajectories, this means, if the
asymptotic solutions fall inside or outside the potential well
defined by the Helmholtz potential. This is analyzed using the {\it
Gr\"unwald-Letnikov} fractional derivative algorithm
\cite{book:Yang,ALGORITHM:1} for our numerical
simulations, in which we change certain initial condition, the
fractional derivative and the forcing amplitude for two values of
the damping parameter. These two damping values have been chosen in
order to set the system in an underdamped or in an overdamped case.

In fact, the Helmholtz oscillator has a very rich dynamical behavior
for small values of the damping parameter where both, chaotic and
periodic motions can take place and also escapes from the potential
well are possible. On the other hand, when the damping parameter
grows, the system becomes more and more predictable until all
particles fall into the well. However, in the overdamped case, the
effect of the fractional derivative can generate different kind of
motions, including chaotic motion. That is why the underdamped case
has been the most interesting one to study in the nonlinear dynamics
field in the last decades, but the overdamped case can be
interesting when fractional derivatives are involved.

By using the above algorithm we calculate the orbits in phase space
and show, in the parameter space, the attractor and the escape
times, i.e., the final state of the trajectories, inside or outside
the potential well, and the time to reach such state, respectively.
The parameters that we change in the simulations are: the damping
term $\mu$, the fractional derivative $\alpha$, the forcing $F$ and
the initial condition $x_0$. For
visualization purposes, we have plotted in parameter space, both
the attractor and the escape time gradient (as can be seen along the
manuscript). In these figures, it is possible to appreciate the
complexity of the relation between the fractional parameter and the
dynamics of the system. In fact, the attractor boundaries show
some sort of fractalization, which means that a small deviation in the parameters
value can lead to drastically different solutions, i.e., the system
asymptotic behavior can stay inside the well or leave it. It clearly
means that, in the fractal regions, errors in the parameters might cause
different final states, as is well known in chaotic systems.
Indeed, these results show that using the fractional parameter $\alpha$
as a control parameter, can provide different dynamics for the dynamical system, either periodic or chaotic motions and transitions between them. This is precisely one of the main goals of the research work reported here by using phase-space visualizations techniques. Among the goals, we can find a natural extension to the fractional case the well-known results obtained for the non-fractional case in both cases, the
underdamped and the overdamped regimes.

The organization of this paper is as follows. We describe the
fractional damped Helmholtz oscillator, in Sec.~II. The dependence
of the trajectories in function of the fractional parameter $\alpha$
for the underdamped case is carried out in Sec.~III. Section IV
shows the distribution of the escape times of the particles from the
potential well versus $\alpha$ and also the analysis of the
bifurcation diagrams in function of $\alpha$ for an overdamped case.
Conclusions and a discussion of the main results of this paper are
presented in Sec.~V.

\section{Fractional Damped Helmholtz Oscillator} \label{sec:fractional_model}

\begin{figure}
\centering
\includegraphics[width=0.55\textwidth,clip=true]{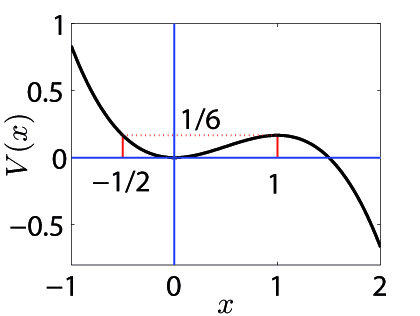}
\caption{Plot of the potential well defined through the function
$V(x) = \frac{1}{2}x^{2} - \frac{1}{3}x^{3}$. Note that for the
interval $x\in[-\frac{1}{2},1)$ in absence of forcing and friction
the orbits are bounded.} \label{fig:1}
\end{figure}

The Helmholtz oscillator is represented by a nonlinear second order
differential equation considering the presence of the potential
defined as $V(x) = \frac{1}{2}x^{2} - \frac{1}{3}x^{3}$, as shown in
Fig.~\ref{fig:1}. It represents the equation of
motion of a unit mass particle in this potential under the influence
of both a periodic forcing and a dissipative force and is given by

\begin{eqnarray}\label{eq:classical_helmholtz}
\ddot{x} + \mu \dot{x} + x - x^2 =  F\cos(\omega t),
\end{eqnarray}
where $\mu$ is the damping parameter, $F$ and $\omega$ the forcing
amplitude, and the forcing frequency, respectively, taking all of
them positive values. Normally, the damping
term, considered like the first derivative in
Eq.~(\ref{eq:classical_helmholtz}), has an impact proportional to
the constant $\mu$ in the system. In what follows, the first
derivative of Eq.~(\ref{eq:classical_helmholtz}) is replaced by an
order $\alpha$ fractional derivative, i.e, $\dot{x}(t)\rightarrow
D^{\alpha}x(t)$, for which we have the \textit{Fractional Helmholtz
Oscillator}. We consider this exchange because of the interest in
looking for a more general dynamics of the oscillator as a
consequence of considering the memory effects through the fractional
operator. As a matter of fact, the most relevant objective here is
to investigate changes over the whole dynamics taking fractional
order derivatives in the interval $[0,2]$ of $\alpha$ values, as is
usually studied in the context of fractional nonlinear oscillators.
Then, the \textit{Fractional Helmholtz Oscillator} with fractional
damping reads

\begin{eqnarray}\label{eq:fractional_helmholtz}
\ddot{x} + \mu D^{\alpha}x + x - x^2 =  F\cos(\omega t),
\end{eqnarray}
with the additive property of the fractional derivative, one can get
a new system for the purpose of our numerical simulations, which can
be expressed as:

\begin{eqnarray}\label{eq:additive_property}
D^{\alpha_1}D^{\alpha_2}x(t) = D^{\alpha_1 + \alpha_2}x(t)
\end{eqnarray}

Thus, the new system is composed by a set of three fractional
differential equations according to Eq.~(\ref{eq:additive_property})
and they read as follows:

\begin{eqnarray}\label{eq:system_helmholtz}
D^{\alpha}x  & =  & y \\ \nonumber D^{1-\alpha}y  & =  & z \\
\nonumber Dz  & = &  F\cos(\omega t) + x^2 - x -\mu y, \nonumber
\end{eqnarray}
where $z$ is a mathematical component coming from the transformation
of the model into a fractional order system. To perform the solution
of Eqs.~(\ref{eq:system_helmholtz}), we use the {\it
Gr\"unwald-Letnikov}~\cite{ALGORITHM:1} fractional derivative, for
which the algorithm to numerically solve this system is given by

\begin{eqnarray}\label{eq:numerical_system_helmholtz}
x(t_{k})  & =  & y(t_{k-1})h^{\alpha} -
\sum_{j=\upsilon}^{k}c_{j}^{(\alpha)}x(t_{k-j}) \\ \nonumber
y(t_{k})  & =  & z(t_{k-1})h^{1-\alpha}  -
\sum_{j=\upsilon}^{k}c_{j}^{(1-\alpha)}y(t_{k-j})\\ \nonumber
z(t_{k})  & = &  \Psi h -
\sum_{j=\upsilon}^{k}c_{j}^{(1)}z(t_{k-j}), \nonumber
\end{eqnarray}
where $\Psi =  F\cos(\omega t_{k}) + x^2(t_{k}) - x(t_{k})-\mu
y(t_{k})$ and $h$ is the discrete time step. The coefficients
$c_{j}^{\alpha}$ are the binomial coefficients derived in the
numerical scheme implemented, $c_{0}^{\alpha}=1$ and

\begin{eqnarray}
c_{j}^{\alpha}=(1-\frac{\alpha + 1}{j})c_{j-1}^{\alpha}.
\end{eqnarray}

In the next subsection, we numerically analyze the dynamics of our
model with the fractional algorithm for $\alpha=1$ and with a
non-fractional algorithm. Then, we compare the results.


In order, and with the only purpose, to test the numerical solutions of the fractional numerical
scheme, a comparison with a fourth order Runge-Kutta method is
carried out. This comparison only is possible for $\mu=0$ or $\alpha=1$ which
corresponds with the non-fractional case.

\begin{figure}
   \centering
   \includegraphics[width=0.95\textwidth,clip=true]{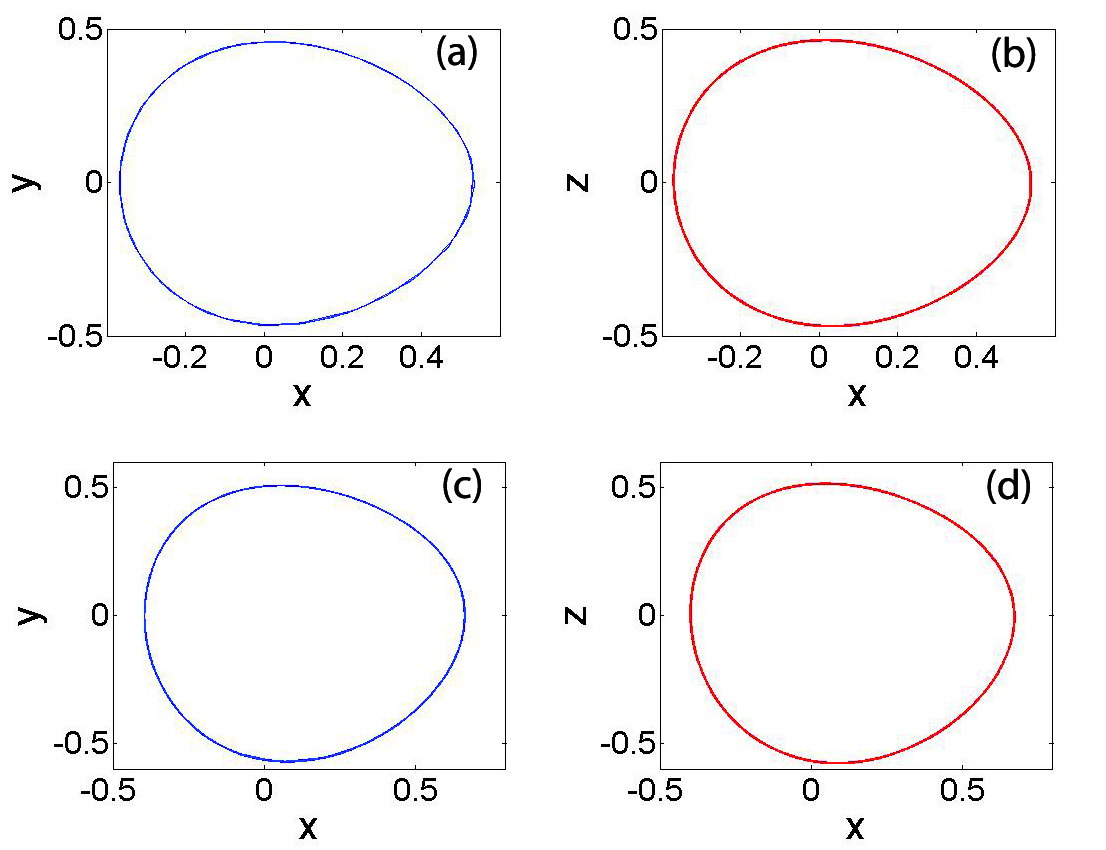}
\caption{Comparison between solutions provided by distinct methods
for the damped Helmholtz oscillator at the zero initial conditions
with  $\mu = 0.1$ and $F=0.1$ in panels (a) and (b), $\mu = 0.8$ and
$F=0.46$ in panels (c) and (d). Panels (a) and (c) show the
trajectories of the solution of the fourth-order Runge-Kutta scheme.
Panels (b) and (d) show the trajectories of the solution of the
fractional numerical scheme with $\alpha=1$.} \label{fig:2}
\end{figure}

\begin{figure}
   \centering
   \includegraphics[width=0.95\textwidth,clip=true]{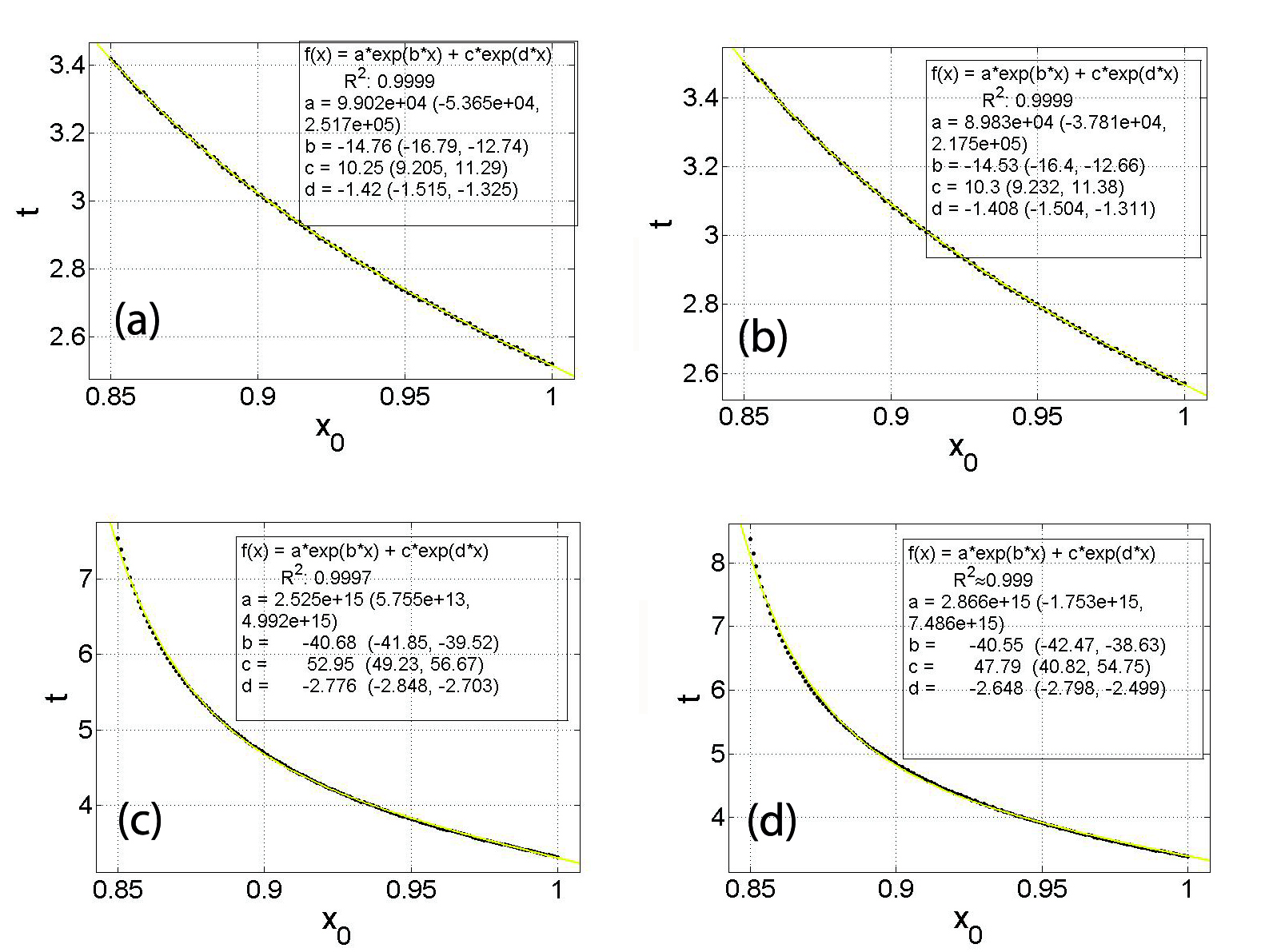}
\caption{Comparison between escape times of solutions provided by
distinct methods for the damped Helmholtz oscillator at $x_0 \in
(0.85, 1)$ with  $\mu = 0.1$ in panels (a) and (b), $\mu = 0.8$ in
panels (c) and (d). In all the panels $F=0.46$ and $\omega = 1$.
Panels (a) and (c) show the escape times calculated with the
Runge-Kutta method. Panels (b) and (d) show the escape times
calculated with the fractional numerical scheme and $\alpha=1$. In
the textboxes we can see the fitting equations and their parameter.
Near the parameter, in the parenthesis, we show the $95\%$
confidence bounds. This is important to check that the fitting
equations and their parameters values of each couple of panels,
(a)-(b) and (c)-(d), are compatible within each other.}
\label{fig:3}
\end{figure}

In Fig.~\ref{fig:2}, we show the resulting orbits starting at zero
initial conditions, which means $ (x_{0}, y_{0}, z_{0}) = (0, 0,
0)$, for both cases the underdamped and the overdamped, the other
parameter values are shown in the caption. In
Figs.~\ref{fig:2}(a)-(c) the trajectories are calculated by using
the Runge-Kutta method and in Figs.~\ref{fig:2}(b)-(d) with the
Gr\"unwald-Letnikov method with $\alpha=1$. In this case, the
fractional method scheme gives us the same trajectories as the one
for the Runge-Kutta method, showing bounded motions of the particle
inside the potential well. This is a first proof of the adequate
response of the fractional algorithm. Now, we focus on an
interesting point that we investigate, the existence of orbits that
escape from the potential well, defined above, and the time that the
particles need to leave the potential well, namely the escape time.
In Fig.~\ref{fig:3} we represent an extra comparison between the two
methods by plotting the escape times of the orbits varying the
initial conditions inside the well, for parameter values for which
the particles escape from the well. For that purpose, we set
$F=0.46$ and $\omega=1$ with $\mu=0.1$ in Figs.~\ref{fig:3}(a)-(b)
and with $\mu=0.8$ in Figs.~\ref{fig:3}(c)-(d). Then, we vary the
initial conditions in the interval $x\in (0.85, 1)$. In the
non-fractional Helmholtz oscillator, the effects of dissipation help
to keep the particles inside the well. For that reason, when the
damping term grows bigger and the system is in an overdamped regime,
the dynamics of the system becomes more predictable since for all
initial conditions the particles need more time to escape the well
and more initial conditions fall into the bottom of the potential
well. On the other hand, in the case of weak dissipation, the
dynamics of the system are very rich and the escape times become
smaller. In fact, in the panel it is possible to appreciate that the
escape time follow that trend, since for $\mu=0.8$ the particle
needs more time to leave the potential well. As a summary of our
comparison, also here, in the left figures of the panel we have used
the Runge-Kutta integrator, while for the others the fractional
numerical scheme. In general, the orbits take more time to escape in
the fractional scheme, as supported by Fig.~\ref{fig:3}, due to the
characteristics of the Gr\"unwald-Letnikov method of
integration~\cite{Scherer}. However, for the sake of the results of
this paper, we want to stress out that, in both numerical schemes,
the escape time curves follow the same trend. This last affirmation
can be proved trough the comparison of the fitting curves, that
appear in the textboxes of each figure. Starting from
Figs.~\ref{fig:3}(a)-(b), we can see that the fitting curve is
exponential for both algorithms, but also that the parameter values
of the first one falls within the $95\%$ confidence bounds, that are
the numbers inside the parenthesis of the other one, and vice versa.
The same thing happens for the fitting curves in
Figs.~\ref{fig:3}(c)-(d). Therefore, we can confidently affirm that,
in both cases, the fitting curves of the escape times calculated
with the Runge-Kutta method follow the same equation of the ones
calculated with the Gr\"unwald-Letnikov algorithm and their
parameter match with a confidence at least of the $95\%$. This gives
us another good proof of the reliability of the fractional
algorithm. That settled, we can start our analysis of the system.

\section{Dynamics of the Fractional Helmholtz Oscillator in the underdamped case} \label{sec:results_1}
Here, we show numerical simulations in order to understand the
dynamics of the fractional Helmholtz oscillator involving a
fractional order damping for different values of the damping
parameter $\mu$. We fix the parameters of the system for the
underdamped case as $\mu = 0.1$, $F=0.1$ and $\omega = 1$ which are
a very convenient choice for this study. In this physical situation,
as previously mentioned, the dynamics of the non-fractional
Helmholtz oscillator is very rich. In this case, the fractional
damping term will add more complexity in both the dynamics and the
topology of the system as we show now.

At first, we focus our attention to the final state of the system,
whether the trajectories stay inside the potential or leave it.
Then, we study the variation of the escape times of the particles
from the well. The escape time, $t$, is the time that the particle
spends inside the well before crossing the boundary ($x=1$ in this
model) escaping from it. In fact, we consider that the particle has
escaped when it crosses the point $x>1$ with a positive velocity,
$\dot{x}>0$. For that purpose, we start with the analysis of the
fractional dynamics by changing the $\alpha$ and the $x_0$ initial
condition values and find critical values where the behavior of the
system changes abruptly. This analysis keep the other initial
condition equal to zero and the same values of the rest of the
constant parameters as previously stated.
\begin{figure}
   \centering
   \includegraphics[width=0.95\textwidth,clip=true]{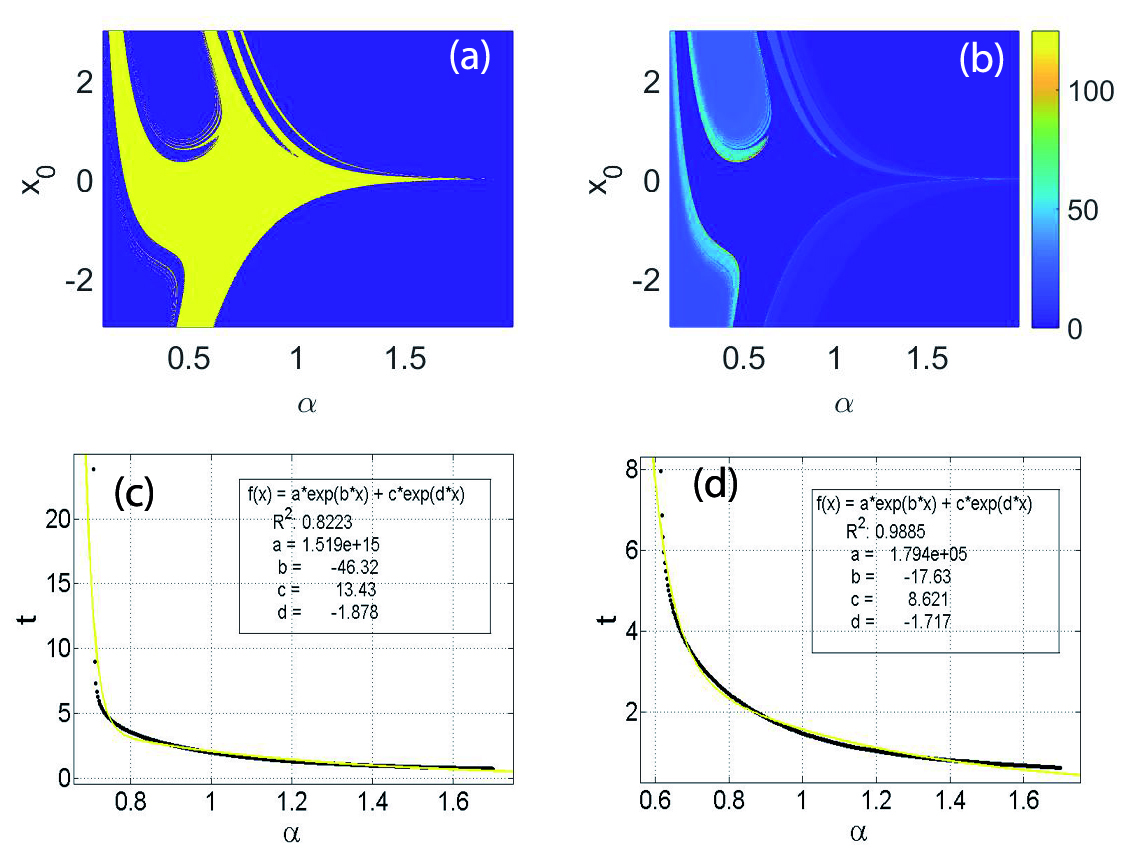}
\caption{Figure (a) shows the final state of the system depending on
the value of $\alpha$, from $0.1$ to $2$, and the variation of the
first initial condition $x_0$ from $-3$ to $3$. White color (yellow
on line) implies that the particle remains inside the well and black
color (blue on line) that particle escapes. In (a) it is possible to
see the areas where the boundaries show the fractalization.  Figure
(b) shows the gradient of the particle escape times from the
potential well depending on the $\alpha$-value and the variation of
the first initial condition for the same values of figure (a).
Figures (c) and (d) show an example of the trend of the escape time
in function of $\alpha$, for different initial conditions, $x_0=-2$
and $x_0=-3$ respectively. } \label{fig:4}
\end{figure}
\begin{figure}
   \centering
   \includegraphics[width=0.95\textwidth,clip=true]{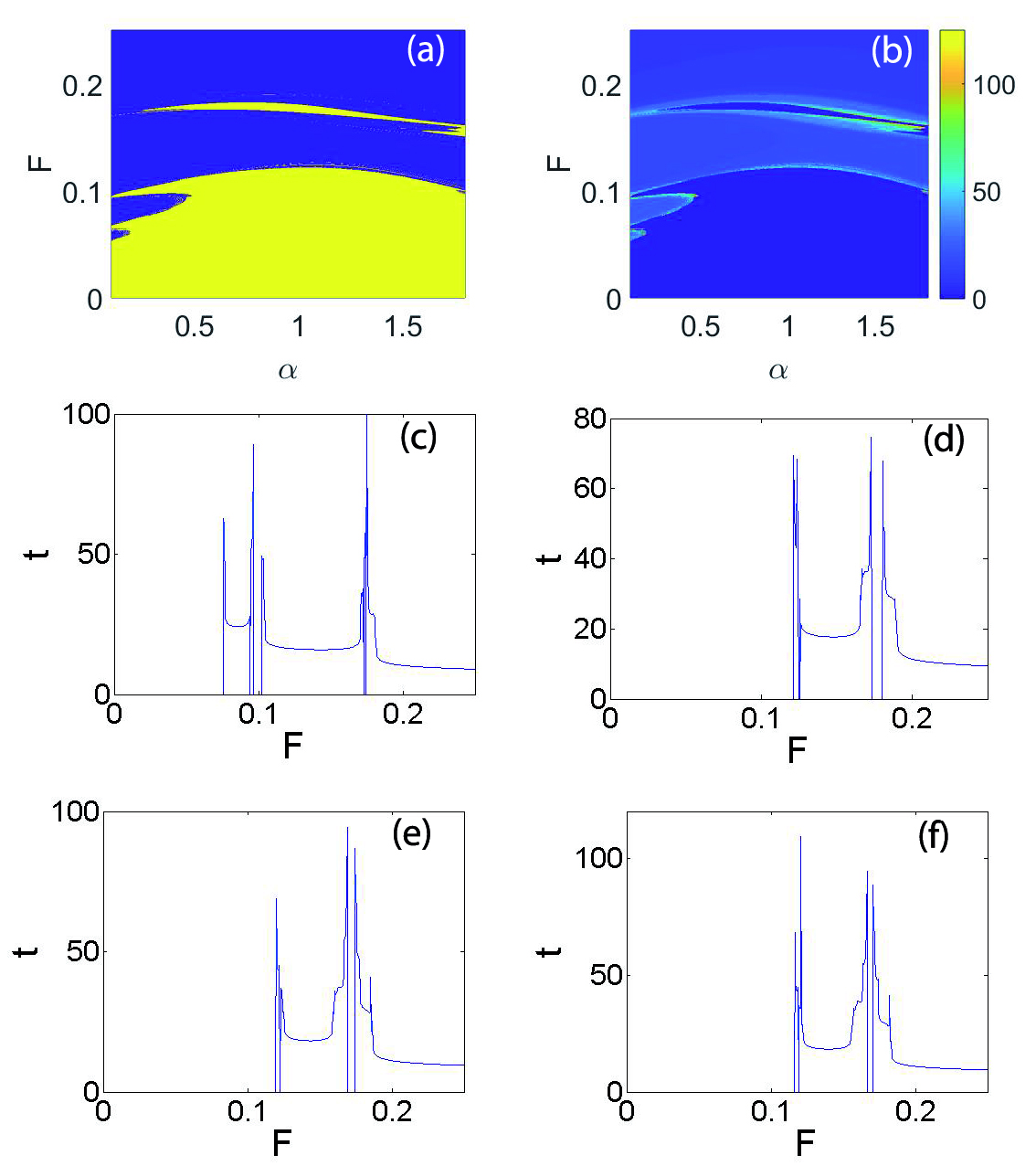}
\caption{Figure (a) shows the final state of the system taking
$\alpha =[0.1,1.8]$ and $F=[0, 0.3]$ and keeping constant the rest
of the parameters. Gray color (yellow on line) implies the particle
that remains inside the potential well and black color (blue on
line) that particle escapes from it. Figure (b) shows, for the same
range of $F$ and $\alpha$ and the same parameter as in figure (a),
the gradient escape times. The other figures show slices of the
figure (b), for different $\alpha$ values, such as $0.25, 1, 1.25,
1.39$, respectively. } \label{fig:5}
\end{figure}
\begin{figure}
   \centering
   \includegraphics[width=0.95\textwidth,clip=true]{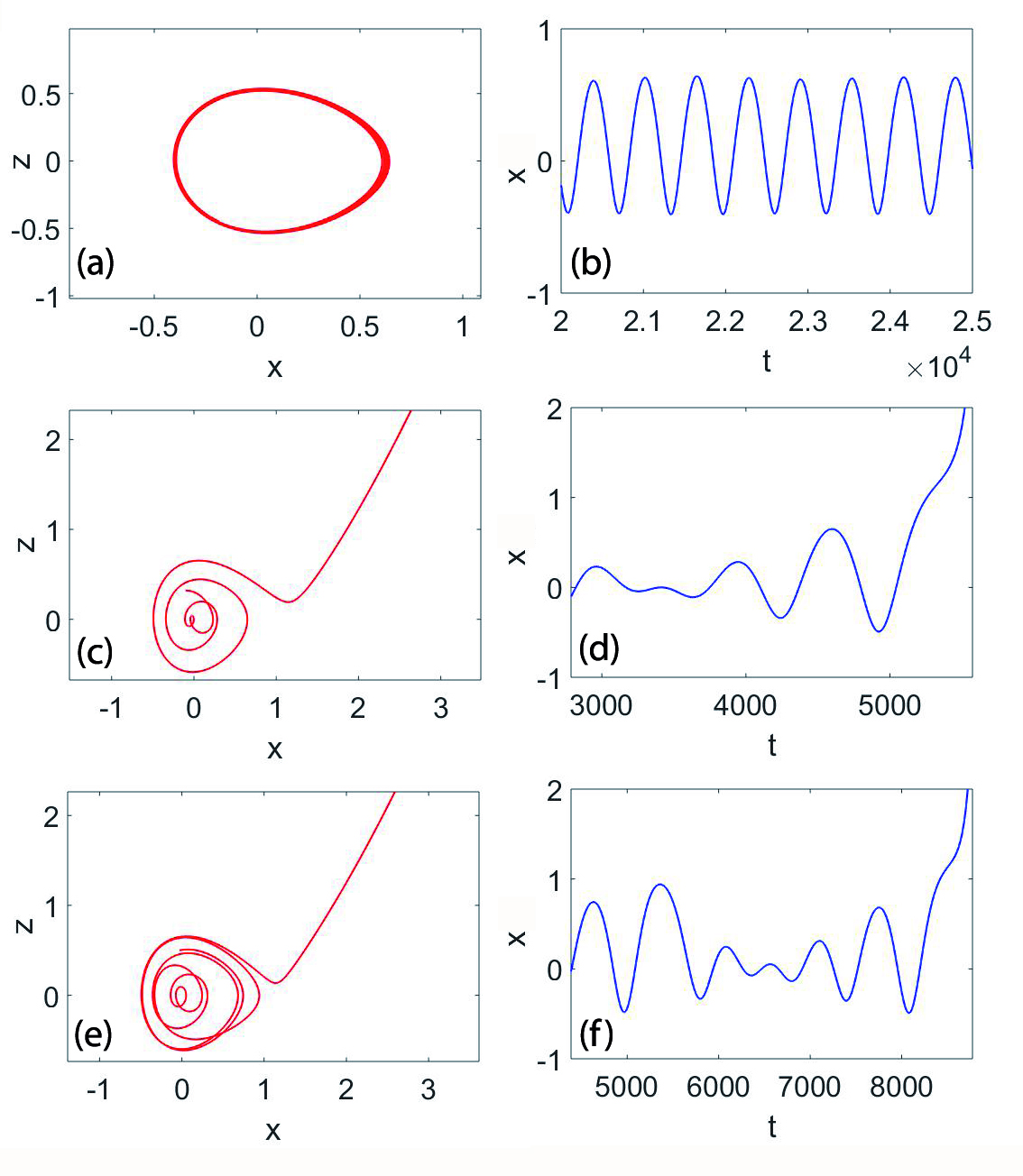}
\caption{Behavior of the fractional damping Helmholtz oscillator
with $\mu=0.1$, $F=0.1$ and $\omega = 1$. The figures represent the
trajectories of the Helmholtz oscillator $z$ vs $x$ and the
oscillations of the position $x$ in time with different $\alpha$
values at different initial conditions $x_0$. In particular
$\alpha=0.39$, $x_0=0.4$ (a) and (b), $\alpha=0.4$, $x_0=0.4$ (c)
and (d), $\alpha=0.41$, $x_0=0.5$ (e) and (f). In figures (a) and
(b) the plots represent the typical bounded orbit inside the
potential well. In figures (c) and (d) the plots show a faster
escaping trajectory. In figures (e) and (f) the plots represent a
slower escaping trajectory. } \label{fig:6}
\end{figure}
Thus, we can see in Fig.~\ref{fig:4}(a) the final state of the
system by varying the values of the factional derivative $\alpha$
and the initial condition $x_0$. That means that we have plotted the
attractors, being an attractor a set of numerical values toward
which a system tends to evolve, for a wide variety of starting
conditions of the system. So that, it is possible to appreciate two
recognizable attractors, the gray one (yellow online) that represent
the case when the particle remains inside the well and black one
(blue online) that represent when the particles escape. It is, also,
possible to appreciate the fractalization of the boundaries between
the different region of the figure. In Fig.~\ref{fig:4}(b) we show
the gradient of the escape time. Here, it is possible to appreciate
that near the boundaries that separate the two attractors defined in
Fig.~\ref{fig:4}(a) the escape times are larger and, in particular,
in the zones that show fractalization. Also, we depicted the trend
of the escape times in function of $\alpha$ to visualize how they
change along the boundaries of the attractor. In order to avoid
fluctuations in the  escape times, we decided to move along the
smoother boundary of the attractor, i.e., for $x_0<0$ and
$\alpha>0.5$, so we have fixed the initial conditions, $x_0=-2$,
Fig.~\ref{fig:4}(c), and $x_0=-3$, Fig.~\ref{fig:4}(d) and varied
$\alpha$. It is possible to see that the two curves do not share the
same trend, as in the first one the exponential curve does not fit
the data very well, $R^2\approx0.8$, while in the second case it
works perfectly, $R^2\approx0.99$. The same thing happens for a
fixed $\alpha$ and varying $x_0$. So, we can say that for this value
of $\mu$ the trend of the escape times depends on the parameter
values, chosen on the boundaries of the attractor. Therefore, it is
not possible to find a common decay law for the escape times, just
that the farther from the attractor the lower the escape time.

Then, to extend our study of the behavior of the system we have also
decided to plot the final state and the escape time gradient in
function of the variation of the $\alpha$ value and the amplitude of
the forcing $F$, for the zero-initial condition, plotted in
Figs.~\ref{fig:5}(a)-(b), respectively. It is interesting to see
that for $\alpha=1$, the particle escapes the potential well for,
approximately, all the forcing amplitude values bigger than
$F\approx0.2$, a well known result \cite{SH:2008}. Moreover, it is
interesting to stress out that $F\approx0.2$ seems to be the maximum
value of the forcing amplitude for which we can have bounded
trajectories independently of the values of $\alpha$. Also, some
details in the figures are interesting, such as the escape region
delimited by $0<\alpha<0.5$ and by $0.05<F<0.1$ and the gap between
$F\approx0.1$ and $F\approx0.17$, for all the values of $\alpha$.
Also in this case, as in Fig.~\ref{fig:4}, the higher escape
times are related with the fractalization of the boundaries between
the different basins of attraction. In order to have a better insight
of Fig.~\ref{fig:5}(b), we have plotted Figs~\ref{fig:5}~(c-f).
These figures have been computed for the zero-initial conditions and different values of the fractional parameter,
$\alpha=0.25$, $\alpha=1$, $\alpha=1.25$ and $\alpha=1.39$,
respectively. Here again, it is not possible to establish a trend
of the escape time in function of $F$ for a fixed $\alpha$ value,
neither it is possible in function of $\alpha$ for a fixed $F$
value, due to the fractalization of the boundaries and the shape of
the basins.

To complete our investigation, we have plotted the Fig.~\ref{fig:6}.
Here we can find three trajectories in the ${x-z}$ plane and the
time series of $x(t)$. In Figs.~\ref{fig:6}(a)-(b) we represent a
bounded trajectory with its related time series. In
Figs.~\ref{fig:6}(c)-(d), we represent an escaping trajectory and
its related time series. In Figs.~\ref{fig:6}(e)-(f) also an
escaping trajectory and its time series with a larger escape time.
This last trajectory has been chosen in a fractal boundary where the
escape time are, in effect, larger.  In the plot of these last
figures, we have chosen the parameter values reported in the
captions of the panel by using the data extrapolated by the
precedent figures, Figs.~\ref{fig:4} and ~\ref{fig:5}, to help
visualize the dynamics of the system while the
parameters, $\alpha, F, x_0$ change and the $\mu$ parameter is
fixed.

\section{Dynamics of the Fractional Helmholtz Oscillator in the overdamped case} \label{sec:results_2}

In this section, we investigate the effect of large values of the
damping parameter $\mu$ on the dynamics of the system. So that, now,
we study the case with a bigger damping value, $\mu=0.8$, the
overdamped case, of the fractional Helmholtz oscillator. Notice
that, as already discussed, in the non-fractional case, the
overdamped situation has not physical interest since almost all
particles with initial condition inside the well are trapped into
the potential well. For that reason, the role of the fractional
parameter $\alpha$ will be crucial to enhance chaotic or regular
motions and also escapes from the potential well.

\begin{figure}
   \centering
   \includegraphics[width=0.95\textwidth,clip=true]{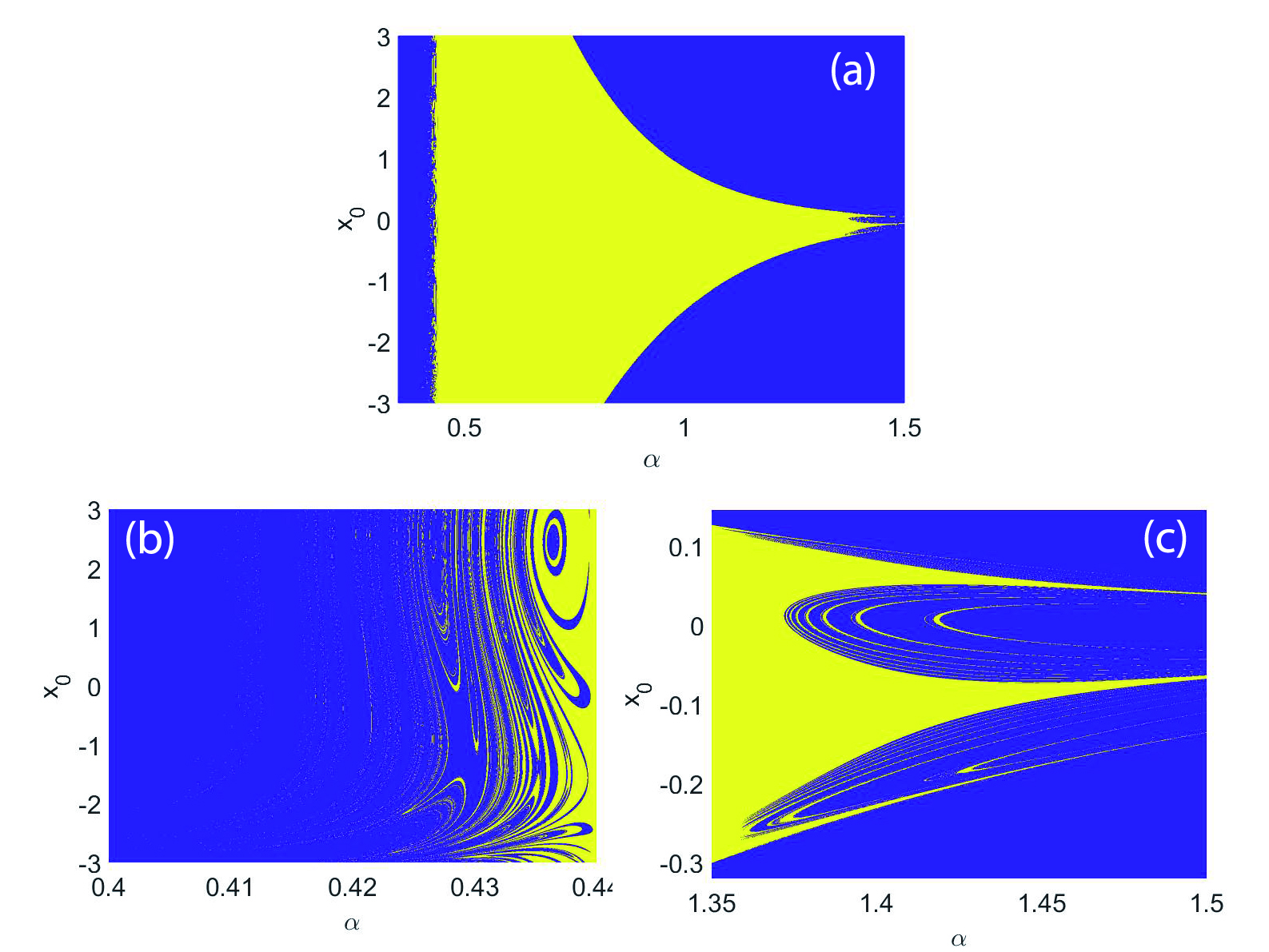}
\caption{Final state of the system depending on the $\alpha$-value
and the variation of the first initial condition $x_0$ from $-3$ to
$3$. White color (yellow on line) implies the particle remains
inside the well and black color (blue on line) that particle
escapes. In figure (a) We can see in this figure a defined area
representing the last state of the particle as time goes on
depending on the initial position $ x_{0} $ and on the $\alpha$
variation, keeping constant the rest of the parameters of the model.
In figures (b) and (c) it is possible to appreciate the zoom of the
figure (a) for values of $\alpha=(0.4, 0.44)$ and $\alpha=(1.35,
1.5)$, respectively. These last ones are the areas where the
boundaries show the fractalization.} \label{fig:7}
\end{figure}

\begin{figure}
   \centering
   \includegraphics[width=0.95\textwidth,clip=true]{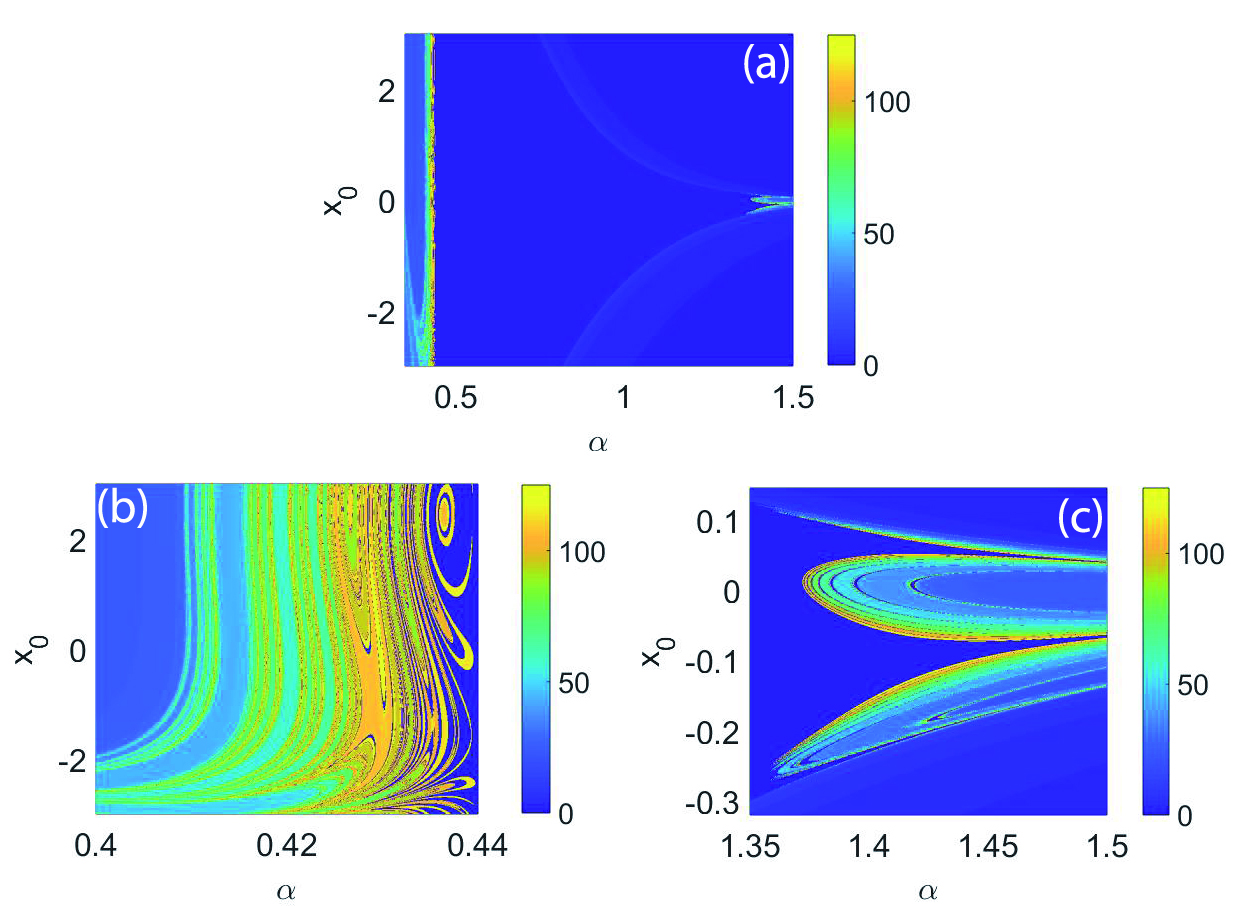}
\caption{Figure (a) shows the gradient of the particle escape times
from the potential well depending on the variation of the $\alpha$
value and of the first initial condition $x_0$ from $-3$ to $3$.
Figures (b) and (c) present a zoom of the most interesting areas of
figure (a), i.e., values of $\alpha=(0.4, 0.44)$ and $\alpha=(1.35,
1.5)$, respectively.  As stated in the previous panel, the areas
where the boundaries show the fractalization.  } \label{fig:8}
\end{figure}

For the reader's convenience, we decide to follow the same path as
in the previous case and started to plot the final state of the
system and the escape times gradient for the simultaneous variation
of $\alpha$ and the first initial conditions $x_0$, maintaining
fixed the value of the amplitude of the forcing $F=0.46$ and its
frequency $\omega=1$. Notice that, with respect to the underdamped
case, while the frequency does not change, the forcing amplitude
that we use is bigger. The results of the computational experiments
are shown in Figs.~\ref{fig:7}(a-c) and Figs.~\ref{fig:8}(a-c). The
first panel shows the final state of the system with the purpose to
know whether the particle remains inside the well or escapes from
it. Gray color (yellow online) has been used to represent the case
when the particle remains inside the well and black color (blue on
line) when it escapes. It is possible to appreciate the
fractalization of the basins in Figs.~\ref{fig:7}(b)-(c). In
particular, in the second one we can see that the fractalization
starts for $\alpha\approx1.36$.

Then, Figs.~\ref{fig:8}(a-c) show the gradient of the escape time.
Here, it is possible to appreciate that, as in the previous case,
near the boundaries defining the non-escaping and escaping regions
the escape times are larger and in particular, in the zones that
show fractalization.

\begin{figure}
   \centering
   \includegraphics[width=0.5\textwidth,clip=true]{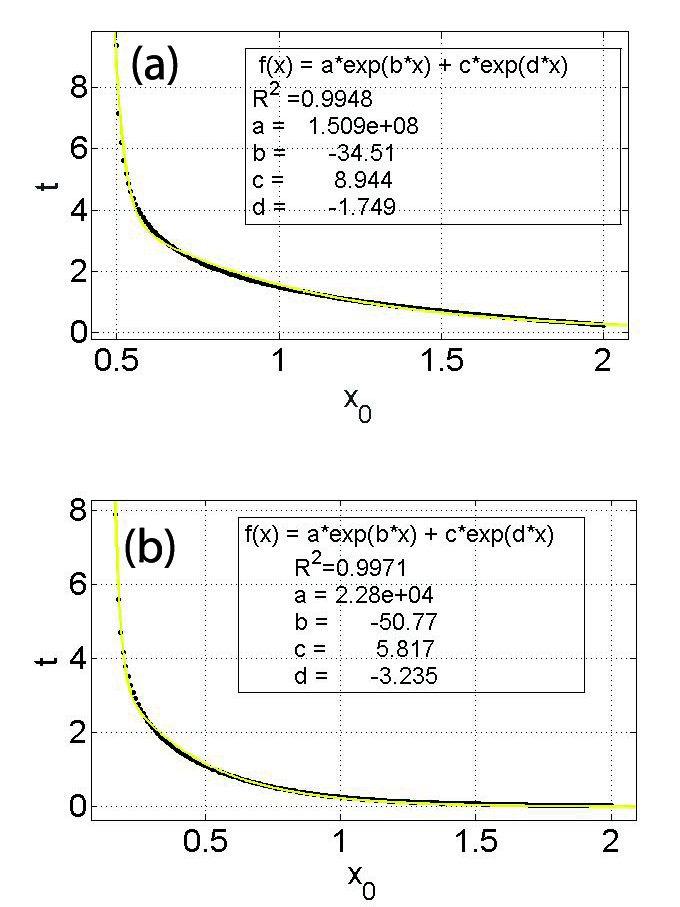}
\caption{(a) Average escape times from the well for $\alpha=1.1$
versus different initial conditions inside the well. (b) Represents
the same for $\alpha=1.3$. It is possible to observe the exponential
decay law.} \label{fig:9}
\end{figure}

In Figs.~\ref{fig:9}(a)-(b), the escape time of the particle is
shown for fixed $\alpha$ and different initial condition $x_0$. In
Fig.~\ref{fig:9} we have set the value of $\alpha$ smaller than
$\alpha\approx1.36$, for which value we previously have found where
the fractalization of the attractor boundaries begins, as shown in
Fig~\ref{fig:7}(c). So that, as in the underdamped case, we use the
existence of smooth boundaries and focus our study on finding a
general decay law for the escape time in function of the variation
of the initial condition $x_0$ and the $\alpha$ parameter. Thus, we
analyze the effects of the higher damping coefficient on the escape
time near the attractor boundaries. We started, in
Figs~\ref{fig:9}(a) and \ref{fig:9}(b), by varying the initial
conditions and fixing the $\alpha$ parameter at $\alpha=1.1$ and
$\alpha=1.3$, respectively. It is possible to appreciate in both
figures the exponential decay law for the time escape in function of
the initial condition $x_0$ where the best fitting is
$t(x_0)=ae^{bx_0}+ce^{dx_0}$. For that reason, we have decided to
explore the variation of the exponential parameters in the decay law
in function of the $\alpha$ parameter.

\begin{figure}
   \centering
   \includegraphics[width=0.5\textwidth,clip=true]{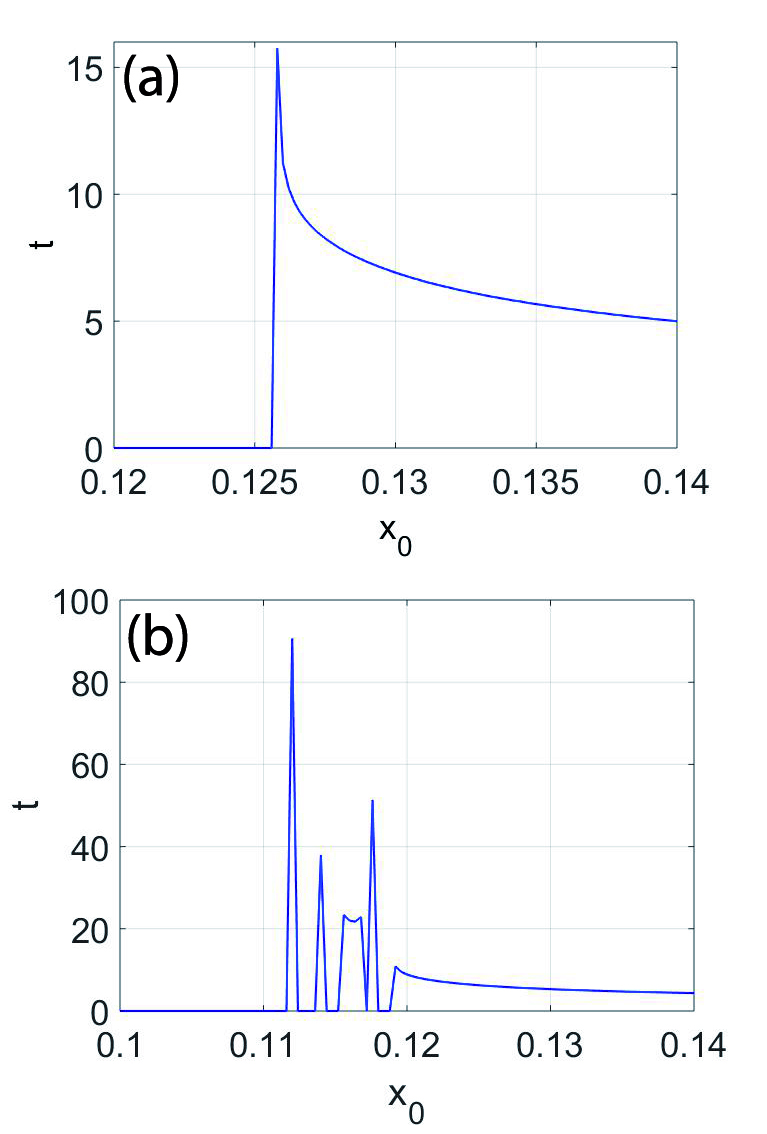}
\caption{(a) Average escape times from the well for $\alpha=1.35$
and different initial conditions inside the well. (b) The same for
$\alpha=1.36$. It is possible to observe that from values of
$\alpha\approx1.36$ the escape times, before to show the monotonic
decay, have fluctuations.} \label{fig:10}
\end{figure}

First of all, we found for $\alpha = 1.36$, some fluctuations in the
escape time before the monotonic exponential decay, as shown in
Fig.~\ref{fig:10}. In this last panel, we show the change in the
behavior of the escape time of the system, when the $\alpha$ value
passes from $1.35$ (Fig.~\ref{fig:10}(a)) to $1.36$
(Fig.~\ref{fig:10}(b)).Through the figures analysis, we can
determine that the appearance of the peaks (Fig.~\ref{fig:10}(b)),
due to the fractalization of the attractor
boundaries~\cite{Scherer}, inhibited us to keep going on to larger
values of $\alpha$ then $1.36$ for the study of the exponential
parameter variation.

\begin{figure}
   \centering
   \includegraphics[width=0.95\textwidth,clip=true]{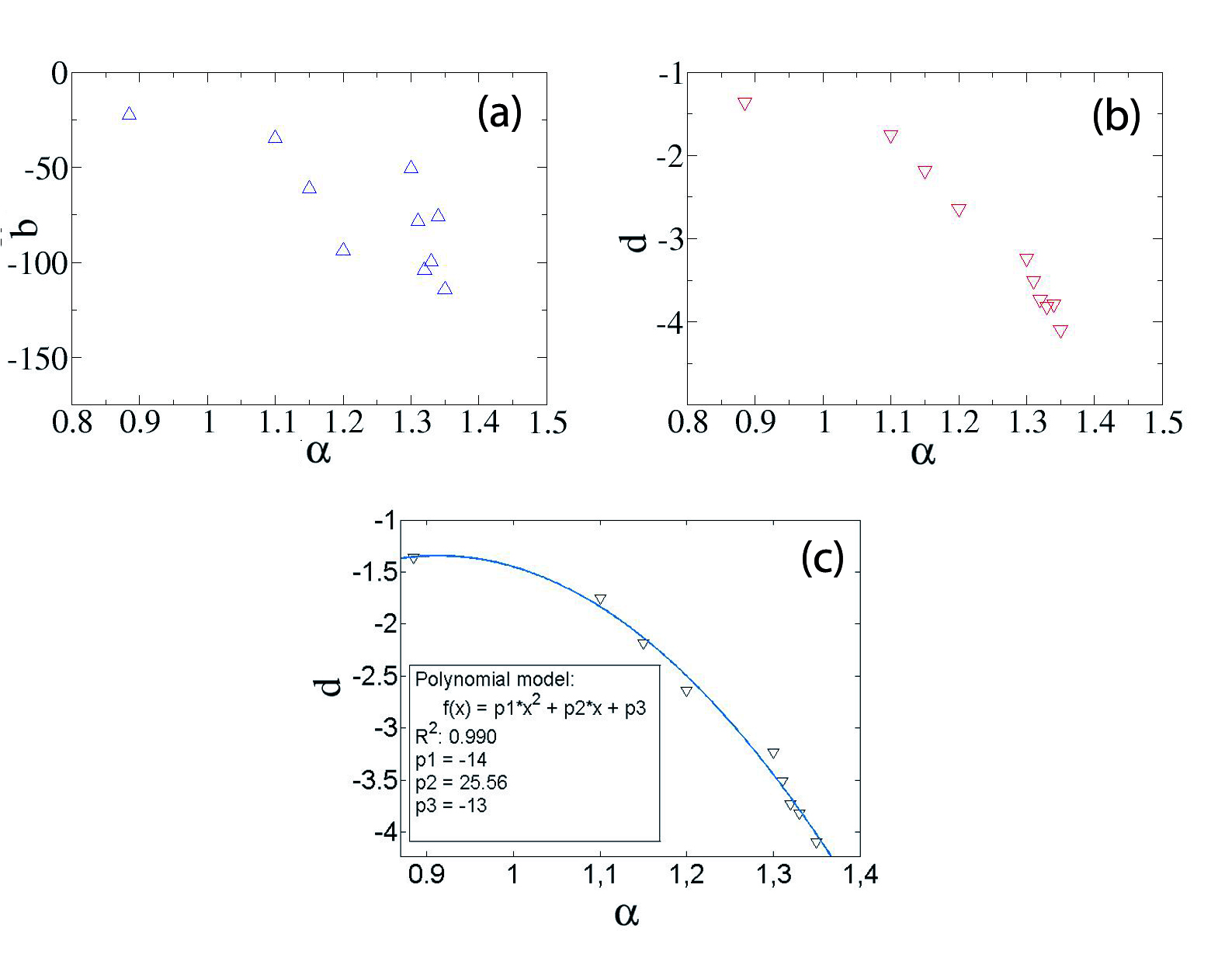}
\caption{The figures (a) and (b) show, respectively, the dynamics of
the decay law exponentials $(b,d)$, calculated as in
figures~\ref{fig:9} in function of the $\alpha$ values smaller than
$\alpha=1.36$, value for which the boundaries fluctuations of the
escape times begin. In figure~(c) a regression curve for the
dynamics of parameter $d$ in function of the fractional derivative
parameter $\alpha$, shown in figure~(b), is proposed.}
\label{fig:11}
\end{figure}

The plot of the exponential parameter $b,d$ in function of the
parameter $\alpha$ can be found in Figs.~\ref{fig:11}(a)-(b),
respectively. It is important to say that, all the curves from which
we have got those values of $b$ and $d$ show a $R^2>0.99$.
Differently from the underdamped case, we can find here a common
trend for the escape time distribution all along the attractor
boundary, possibly an effect of the high damping parameter value. In
Fig.~\ref{fig:11}(a), we can observe some fluctuations for
$0.8<\alpha<1.4$. However, in Fig.~\ref{fig:11}(b), a nonlinear
decrease of $d$ versus $\alpha$ is observed. The corresponding terms
of this decay law are depicted in Fig.~\ref{fig:11}(c). In that
figure, a polynomial decay law fits the data very well since the
regression coefficient $R^2\approx0.99$. Therefore, we can write the
mathematical expression of this curve as follows:
\begin{equation}\label{eq:fit_d_a}
  d(\alpha)=p_1\alpha^2+p_2\alpha+p_3,
\end{equation}
with $p_1=-14$, $p_2=25.56$ and $p_3=-13$.

\begin{figure}
   \centering
   \includegraphics[width=0.55\textwidth,clip=true]{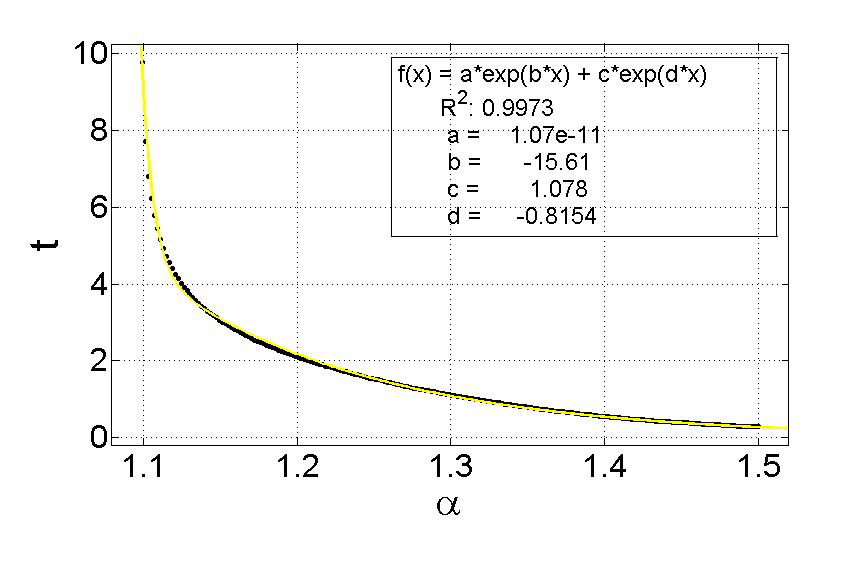}
\caption{Average escape times from the well for fixed $x_0=0.5$ and
from the first $\alpha$ value for which the trajectories escape from
the well, from $\alpha\approx1.0995$,to $1.5$. } \label{fig:12}
\end{figure}

The polynomial fitting of $d$ versus $\alpha$ connects with
the exponential fitting shown in Fig.~\ref{fig:9} and in
Fig.~\ref{fig:12}. For that reason, we can say that the two figures
are related. In fact, the curves in Fig.~\ref{fig:9}(b) and in
Fig.~\ref{fig:12} cross each other in the following points: $t=1.1$
for $x_0=0.5$ in the first one and for $\alpha=1.3$ in the second
one. Furthermore, the trend of the parameter $d$ versus $\alpha$ is
as we expected since the curve of the distribution of the
escape time $t$ in function of $x_0$, reads:
\begin{equation}\label{eq:fit1}
  t(x_0)=ae^{bx_0}+ce^{dx_0}
\end{equation}
in which the first term is negligible because of the parameter
values $a,b$. Therefore, the main term of the regression curve can
be written as follows:
\begin{equation}\label{eq:fit1sim}
  t(x_0)\approx ce^{dx_0}.
\end{equation}
Finally, we substitute $d$ as Eq.~(\ref{eq:fit_d_a}), for values of
the initial condition $x_0>0$ obtaining an exponential decay law
similar to the one shown in Fig.~\ref{fig:12}.

Finally, in the latter figure, the study of the average escape times, by
varying $\alpha$ for a fixed value $x_0=0.5$, shows that the
fractional damping term can be used as a control parameter for the escape
of the particles from the potential well. In this sense, when we
move from points which are on the smooth boundaries of the attractor,
the escape times of the particles decrease insofar the values of
$\alpha$ are going far away from the attractor, as shown in
Fig.~\ref{fig:12}. It can be seen that all of the escape times cases
exposed are contained in Figs.~\ref{fig:7} and \ref{fig:8}.

\begin{figure}
   \centering
   \includegraphics[width=0.95\textwidth,clip=true]{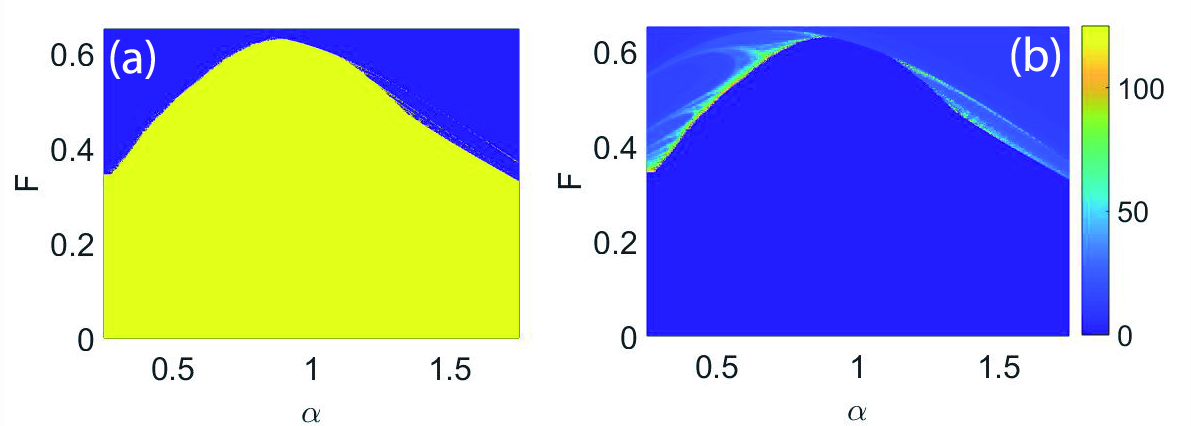}
\caption{In figure (a) the panel shows the final state of the system
taking $\alpha =[0.35,1.5]$ and $F=[0, 0.7]$ and keeping constant
the rest of the parameters. Gray color (yellow on line) implies the
particle remains inside the potential well and black color (blue on
line) that particle escapes from it. In figure (b) the panel shows,
for the same range of $F$ and $\alpha$ and the same parameter as in
figure (a), the gradient escape times.  A semi parabolic shape
denotes the behaviour of the particle as $\alpha$ and $F$ vary.}
\label{fig:13}
\end{figure}

Now, in order to follow the same path as in the underdamped case, we
show the final state of the system in Fig.~\ref{fig:13}(a), and the
escape times gradient in Fig.~\ref{fig:13}(b), in function of the
simultaneous variations of $F$ and $\alpha$. In
Fig.~\ref{fig:13}(a), the parameter values for which the particles
do not escape are colored in gray (yellow online) and the black
color (blue online) is used for the different case. It is
interesting the semi parabolic shape in both
Figs.~\ref{fig:13}(a)-(b). Again, as shown in Fig.~\ref{fig:13}(b),
the closer the values of $F$ and $\alpha$ are to the values on the
boundaries for which there is no escape, the higher the time of the
escape.

\begin{figure}
   \centering
   \includegraphics[width=0.75\textwidth,clip=true]{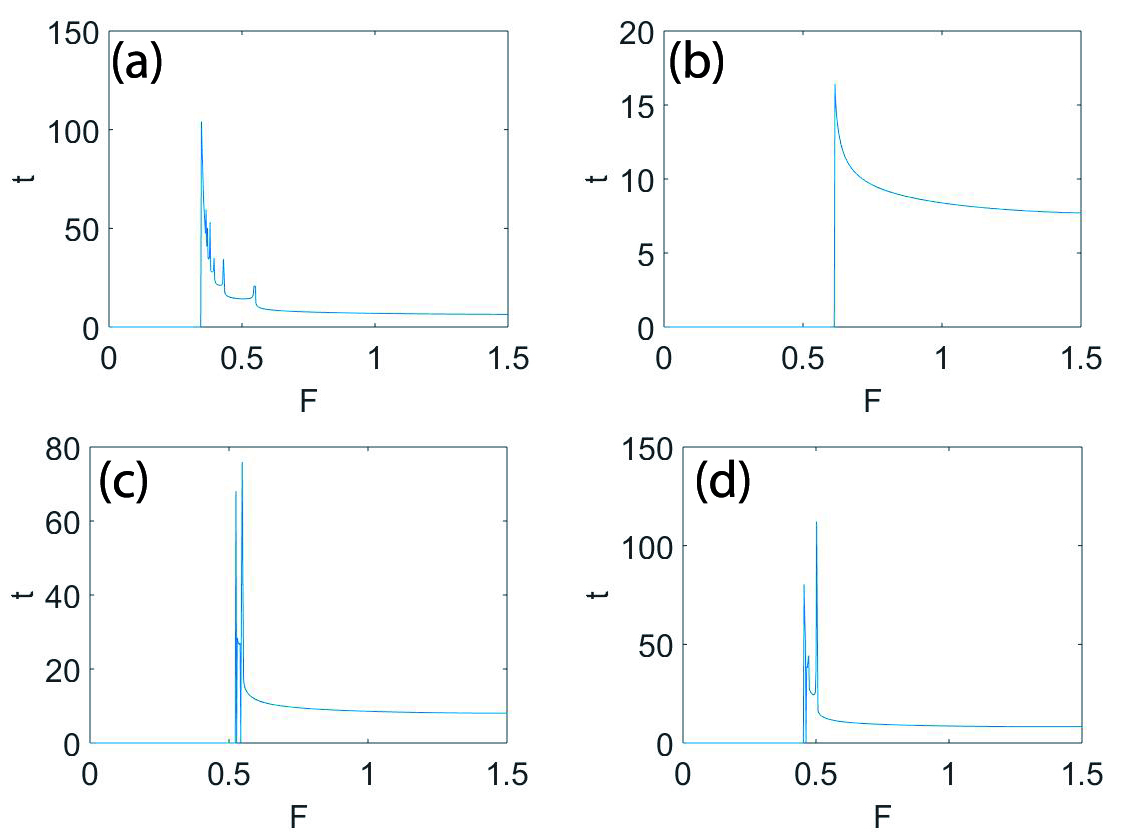}
\caption{The panel shows the particle escape times from the
potential well with $\alpha= 0.25, 1, 1.25, 1.39$, by reading the
figures from (a) to (d), and varying $F$. } \label{fig:14}
\end{figure}

Then, we considered distributions of the escape times, $t$, in
function of the periodic forcing $F$ to study the behavior of the
escape times from another point of view. Different results have been
obtained and are shown in Figs.~\ref{fig:14}(a-d). For the numerical
simulations, we choose the values of $\alpha = 0.5, 1, 1.25, 1.39$,
respectively, and we vary $F$ for the zero-initial conditions.
Figure~\ref{fig:14}(a) represents the case when $\alpha=0.25$. In
this first plot we can see the fluctuations of the escape time for
the first values in the interval, but from $F\approx0.6$ we obtained
a smooth curve due to the strong effect of the high values of $F$.
In the next plot, we have repeated the same experiment with
$\alpha=1$ and we show the results in Fig.~\ref{fig:14}(b). For this
$\alpha$ values, differently from the underdamped case, the figure
illustrates a whole smooth curve for every value of $F$ and, as the
fractalization is not present, very short escape times in comparison
with Fig.~\ref{fig:5}(d). This is a counterintuitive result, as we
have commented along the paper, the higher the dissipation the
longer the escape times. Here, on the other hand, the escape times
diminish in the overdamped case, since the higher value of the
damping term makes the boundaries of the attractors smooth and so
the escape times become smaller. This is a different behavior,
compared with the above case, $\alpha=0.25$. This shows that the
fractalization induced by the fractional term has a deep impact on
the dynamics of the system. The next experiments with $\alpha =
1.25, 1.39$, are respectively shown in Figs.~\ref{fig:14}(c)-(d).
The trend of these two cases is similar. Again, we obtained
fluctuations in the escape times for the first values of $F$ and
near $F=0.6$ the curves become smooth. Therefore, as we already know
from Fig.~\ref{fig:13}, the first value of $F$ for which the
particle escape depend on the value of $\alpha$, and the dependence
is not linear. As it happened in the underdamped case, the
fluctuations, due to the fractalization of the boundaries, impede us
to study a general decay law for the escape times as we did for the
$\{\alpha-x_0\}$ plane, where a smooth boundary was present. All
this analysis suggests that the variation of the parameter $\alpha$
along with the forcing amplitude $F$ can be responsible of deep
changes in the dynamics of the system, due to the induction of the
fractalization of the parameter space. This means that $\alpha$ is a
suitable control parameter for the behavior of the system.

\begin{figure}
   \centering
   \includegraphics[width=0.75\textwidth,clip=true]{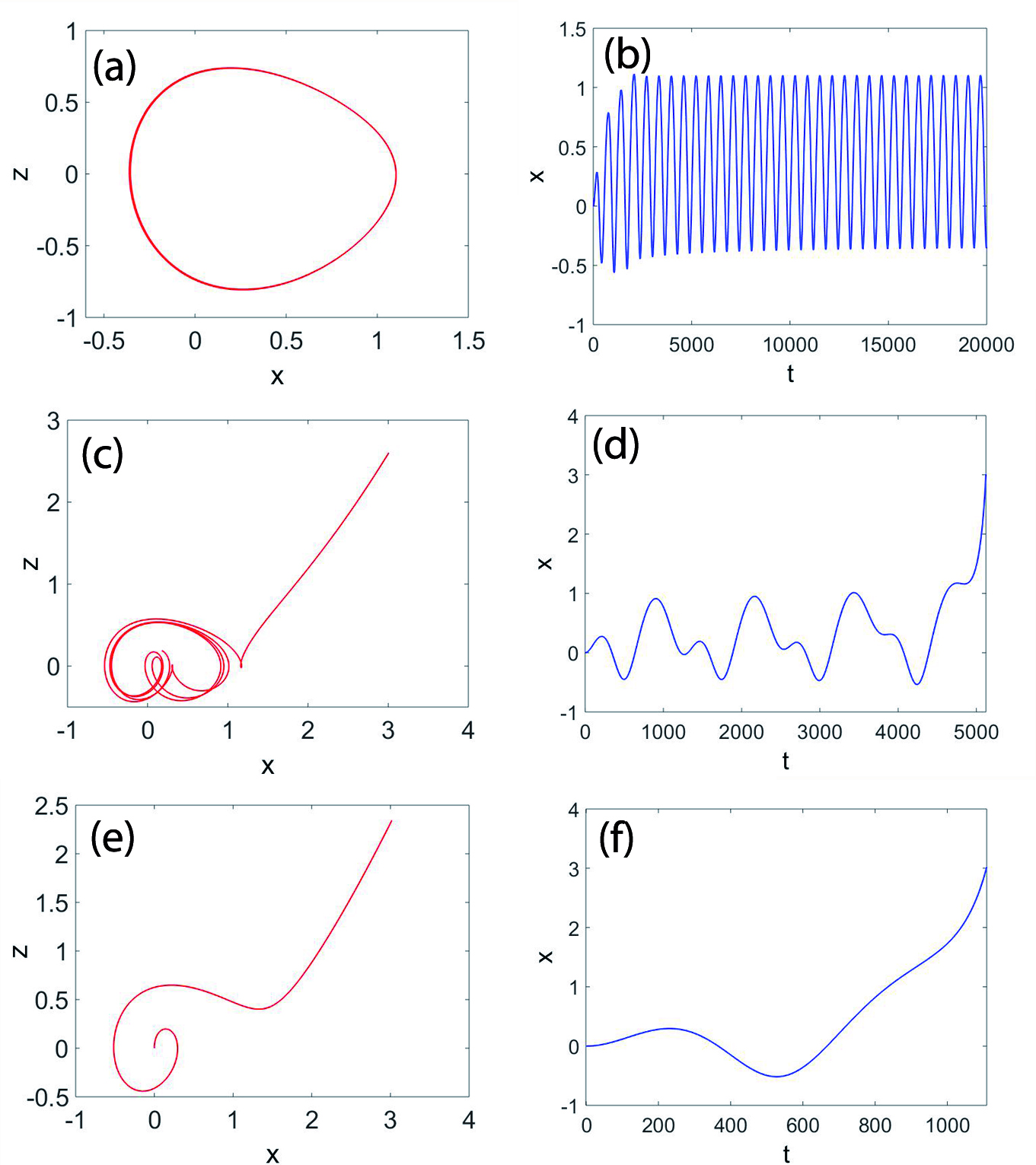}
\caption{The panel shows the behavior of the fractional damping
Helmholtz oscillator with
 different values of $\alpha$ at zero-initial conditions and  $\mu = 0.8$, $F=0.46$
and $\omega = 1$. Figures~(a), (c) and (d)  show the trajectories of
Helmholtz oscillator  $x$ vs $z$, while figures (b), (d) and (f) the
oscillations of the position $x$ with time, for $\alpha=0.5$,
$\alpha=1.39$, $\alpha=1.75$, respectively. Figures (a) and (b)
represents the typical bounded orbit inside the potential well.
Figures (c) an (d) plot the trajectory that escape the potential
well. The last figures plot a trajectory with a shorter escape
time.} \label{fig:15}
\end{figure}
In order to have a better understanding of the
dynamics of the system, we plot the trajectories with different
values of the $\alpha$ parameter in Fig.~\ref{fig:15} by fixing the
value of the damping $\mu=0.8$, the forcing amplitude $F=0.46$, the
frequency $\omega=1$ and the zero-initial conditions. In
Figs.~\ref{fig:15}(a)-(b) we take $\alpha=0.5$ and we can see the
closed orbit and the evolution with time of the trajectory described
by the system. Next case we consider is $\alpha = 1.39$ which is
depicted in Figs.~\ref{fig:15}(c)-(d). This value of $\alpha$
generates escapes from the potential well even considering
zero-initial conditions. This abrupt change in the dynamics is an
interesting issue that reveals, again, the dependence on the
fractional damping term. Another important fact is that this result
confirms the importance to study the effect of taking a specific
value of $\alpha$ and measure the escape times depending on
different initial conditions. Thus, Fig.~\ref{fig:15}(c) shows that
the trajectory is initially contained in the well but at certain
time it escapes from it. In fact, in Fig.~\ref{fig:15}(d), we show
the escaping from the potential well other than the oscillations
before. We also consider the case of $\alpha = 1.75$ another case
for which the trajectory is escaping from the well, according to the
numerical results shown in Figs.~\ref{fig:15}(e) and
\ref{fig:11}(f). The difference is that for this value of $\alpha$
the oscillator behaviors before the escape die out faster than the
case for $\alpha = 1.39$. So that, indeed $\alpha$ can be considered
as a control parameter of the system, which confirms its relevance
in this work.

\begin{figure}
   \centering
   \includegraphics[width=0.95\textwidth,clip=true]{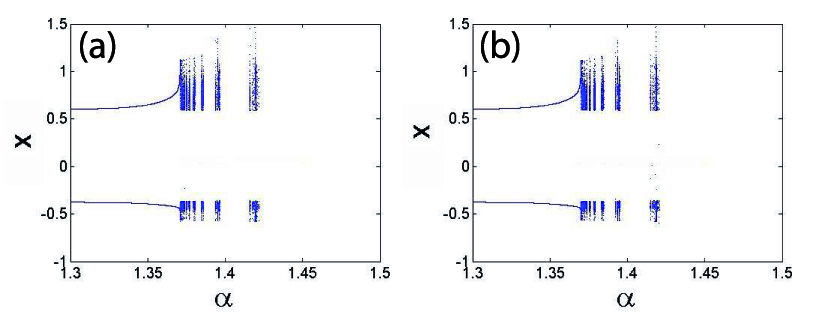}
\caption{Bifurcaction diagram with $x_0 = 0.0$, $\mu = 0.8$,
$F=0.46$ and $\omega = 1$(a). Bifurcaction diagram with $x_0 =
0.01$, $\mu = 0.8$, $F=0.46$ and $\omega = 1$(b)} \label{fig:16}
\end{figure}

Now, to stress out this crucial role of the $\alpha$ parameter, in
the overdamped case, we have considered important to analyze the
dynamics of the system in function of the $\alpha$ variation with a
bifurcation diagram, see Figs.~\ref{fig:16}(a)-(b). These figures
show two different bifurcation diagrams with just a little
difference on their initial conditions. In both figures the dynamics
related with escapes have been left white, for an easier reading of
the graphics. Figure~\ref{fig:16}(a) corresponds to initial
conditions $ (x_{0}, y_{0}, z_{0}) = (0 , 0, 0)$  and  $\alpha$ has
been considered from $1.3$ to $1.5$. Figure~\ref{fig:16}(b) has been
drawn with a slightly different initial conditions, $(x_{0}, y_{0},
z_{0}) =(0.01, 0, 0)$. The figures show the effectiveness of
$\alpha$ as a control parameter. In fact, the results, that match
with the previous ones of Fig.~\ref{fig:7} and Fig.~\ref{fig:8},
confirm that the $\alpha$ parameter can induce chaos in the system,
even in the overdamped case, and that its appearance is robust for
small variation of the initial conditions.

\section{Conclusions} \label{sec:conclusions}
We have studied, using the Gr\"{u}nwald-Letnikov integrator, the
dynamics of the fractional Helmholtz oscillator with fractional term
in the underdamped and the overdamped cases. Since the system allows
the particle to escape the potential well, we used the attractor and
the escape time plots as tools to analyze the impact of the
fractional parameter on the dynamics of the system. In the
underdamped case, the dynamics of the system are already rich and
all kind of behavior are possible. However, changing the fractional
parameter has a big impact on the system dynamics. On the other
hand, the second case that is normally not so interesting because
the high dissipation makes the system more predictable, becomes more
interesting when the fractional derivative is introduced. In fact,
in the overdamped case is where the crucial role of the fractional
parameter becomes evident. When we vary the initial conditions and
the fractional parameter, all kind of behavior of the system are
possible, even chaotic. All this variety of the system dynamics
appears, in the attractor and escape time gradient plots, as a
fractalization of the parameter space in function of the variation
of the fractional parameter.  All this makes the fractional
parameter a suitable candidate to control the asymptotic behaviors
of the system. Moreover, in both cases, the escape time plots showed
us that near the boundaries of the attractor they are higher.
However, in the overdamped case, the escape time distribution of the
particles show an exponential-like decay law. Finally, in the
overdamped case we have studied the bifurcation diagrams in function
of $\alpha$ to contrast the previous results. Also, we have seen
that the appearance of chaotic behavior is robust for small changes,
inside certain interval, of the initial condition, and the $\alpha$
parameter.

To summarize, chaotic and periodic regions appearance in the
parameter space depends on the fractional parameter. So that we can
say that it acts as a control parameter of the dynamics of the
Helmholtz oscillator. In fact, the computation of the basins of the
final state and the gradient of the escape times of our model in the
parameter space corroborates the previous conclusions. We expect
these results to be useful for a better understanding of fractional
calculus in chaotic systems since the memory effects are relevant on
their dynamics.

\section{ACKNOWLEDGMENTS}
This work has been supported by the Spanish State Research Agency (AEI) and the European Regional Development Fund (ERDF, EU) under Projects No.~FIS2016-76883-P and No.~PID2019-105554GB-I00, and the National Natural Science Foundation of China (Grant No. 11672325).

\section*{Conflict of Interest:}
The authors declare that they have no conflict of interest.

\end{document}